\documentclass[aps,prd,reprint,groupedaddress,nofootinbib,showpacs]{revtex4-1}

\usepackage{graphicx}
\usepackage{epsfig}
\usepackage{amsmath}
\usepackage{color}
\usepackage[normalem]{ulem}
\usepackage{natbib}

\begin{document}


\title{Towards generating a new supernova equation of state: \\
A systematic analysis of cold hybrid stars
}


\author{Oliver Heinimann}
\email[]{oliver.heinimann@unibas.ch}
\author{Matthias Hempel}
\author{Friedrich-Karl Thielemann}
\affiliation{Department of Physics, University of Basel, Klingelbergstrasse 82,
4056 Basel, Switzerland}


\date{\today}

\begin{abstract}
The hadron-quark phase transition in core-collapse supernovae (CCSNe) has the potential to trigger explosions in otherwise nonexploding models. However, those hybrid supernova equations of state (EOS) shown to trigger an explosion do not support the observational 2~M$_\odot$ neutron star maximum mass constraint. 
In this work, we analyze cold hybrid stars by the means of a systematic parameter scan for the phase transition properties, with the aim to develop a new hybrid supernova EOS. The hadronic phase is described with the state-of-the-art supernova EOS HS(DD2), and quark matter by an EOS with a constant speed of sound (CSS) of $c_{\rm QM}^2=1/3$. We find promising cases which meet the 2~M$_\odot$ criterion and are interesting for CCSN explosions. We show that the very simple CSS EOS is transferable into the well-known thermodynamic bag model, important for future application in CCSN simulations.
In the second part, the occurrence of reconfinement and multiple phase transitions is discussed. 
In the last part, the influence of hyperons in our parameter scan is studied. Including hyperons no change in the general behavior is found, except for overall lower maximum masses. In both cases (with and without hyperons) we find that quark matter with $c_{\rm QM}^2=1/3$ can increase the maximum mass only if reconfinement is suppressed or if quark matter is absolutely stable.
\end{abstract}


\maketitle

\section{Introduction}
Extremely high densities and neutron-rich conditions, which are not directly accessible in terrestrial experiments, are reached in neutron stars. 
With increasing densities the description of matter becomes more uncertain. New degrees of freedom besides nucleons as hyperons and/or quark matter can appear. As an extreme scenario there exists even the possibility of absolutely stable strange quark matter and pure quark stars within Witten's hypothesis \cite{PhysRevD.30.272} (see also earlier works, e.g. \cite{Itoh1970,1978PhRvD..17.1109F}).
The recent discoveries of neutron stars with masses around 2 M$_\odot$ \cite{2010Natur.467.1081D,2013Sci...340..448A,fonseca16}
represent a strong constraint on the appearance and impact of the additional degrees of freedom on the equation of state (EOS). 

In this work, we focus on hybrid stars whose outer parts contain hadronic matter and the inner part quark matter with a first-order phase transition in between. 
In order to systematically analyze hybrid stars in regards to the maximum mass constraint we use the scheme proposed by Alford \textit{et al.} \cite{PhysRevD.88.083013}, applying a simplified, but representative quark EOS. Four different subclasses of hybrid stars were introduced in \cite{PhysRevD.88.083013} according to the stability of hybrid stars at the onset of quark matter and/or the existence of a third family.\footnote{In the mass-radius ($M$-$R$) relation, first family stars are white dwarfs while second family stars are neutron stars. After a phase of instability a third stable branch can build up, which consists of hybrid stars \cite{gerlach68,schertler00}.} In a subsequent work \cite{2016EPJA...52...62A}, a more detailed analysis was presented and different hadronic EOSs were applied. 
Zacchi \textit{et al.}\ \cite{2015PhRvD..92d5022Z,Zacchi16} used the approach of Alford \textit{et al.}\ for comparison of the results obtained with a newly developed SU(3) quark EOS. A special emphasis was put on the occurrence of twin stars, which are pairs of compact stars at equal masses. For the hadronic EOS, they used the relativistic mean-field model DD2 \cite{PhysRevC.81.015803} as we do in the present study.
Alford's classification was also applied in a number of other works \cite{burgio16,ranea16,2015PhRvD..92h3002A,alvarez2016}, varying the hadronic and/or quark EOSs.
Similar parameter scans for quark matter properties were done in \cite{2011ApJ...740L..14W,zdunik13}, where, however, only the maximum mass but not the type of hybrid star was investigated.
One of the main motivations of the present paper is to gather more insights about the parameter space describing the quark matter EOS and the resulting QCD phase transition in the context of core-collapse supernovae (CCSN).

The CCSN explosion mechanism is not yet completely understood. The delayed neutrino-driven (or neutrino-heating) mechanism is the most established and well-investigated one. In one-dimensional simulations no explosions can be obtained except for special low-mass progenitors with an O-Ne-Mg core \cite{2006AA...450..345K}. It has been shown that multidimensional effects such as convection,  nonradial matter flows, or the standing accretion shock instability can trigger a successful explosion. 
One remaining problem is that the resulting explosion energies are typically smaller than the observed values.
Alternative mechanisms such as the acoustic mechanism or the magnetorotational mechanism were also proposed. For more details of the mentioned mechanisms see, e.g., \cite{2012ARNPS..62..407J,2007PhR...442...38J,2015PASA...32....9F,2006RPPh...69..971K}.

Another mechanism showing successful explosions, even in one-dimensional simulations, is the QCD-phase-transition mechanism (see Sagert \textit{et al.}\ \cite{2009PhRvL.102h1101S}). The appearance of quark matter can cause a collapse of the protoneutron star to a more compact configuration, which results in a second shock wave that travels outwards. This second shock wave can revive the stalled first shock and induce the explosion. High explosion energies around and above $10^{51}$~erg \cite{2009PhRvL.102h1101S} make this scenario especially interesting for further investigation. 
However, the hybrid EOSs applied in \cite{2009PhRvL.102h1101S} have maximum masses much below 2~M$_{\odot}$. In the subsequent works exploring this scenario \cite{2009PhRvL.102h1101S,fischer10,2011ApJS..194...39F,fischer12,Fischer2014,2013AA...558A..50N,2014_fischer_polonica}, explosions could not be obtained if the maximum mass was sufficiently high.

On the other hand, only a few SN EOSs that consider quark matter exist (\cite{2009PhRvL.102h1101S,2011ApJS..194...39F,2014_fischer_polonica,Fischer2014,2008PhRvD..77j3006N}), only a few progenitors have been tested, and no systematic evaluation has been done yet. Furthermore, recently it was pointed out that the collapse of the protoneutron star, which was found in the aforementioned works, can be related to the existence of a special third family \cite{2015arXiv151106551H}: 
for the particular EOSs considered, the third family is only marginal at zero temperature, but increases significantly when going to finite entropies as they are found in protoneutron stars. While this points to the importance of the thermal properties of the hybrid EOS, it also implies that a pronounced third family of cold compact stars is favorable for triggering explosions. To which extent this is still possible while being compatible with the 2~M$_\odot$ constraint is one of the main subjects of the present study.

In this paper, we systematically analyze possible parameter configurations of quark matter EOSs. The final aim is to generate a new hybrid SN EOS in the near future that is favorable for explosions and has a ``realistic'' description of the hadronic EOS with good nuclear matter properties. 
We repeat a similar parameter scan as the one of Alford \textit{et al}.\ \cite{PhysRevD.88.083013,2016EPJA...52...62A}. For the hadronic phase, we use the supernova EOS named HS(DD2) \cite{hempel10,Fischer2014}, which is available at finite temperatures and electron fractions and can directly be used in CCSN simulations, and for the quark phase the so-called constant speed of sound (CSS) EOS of Alford \textit{et al}. As a result, we find configurations that support a maximum mass of 2 M$_\odot$ and show a third-family feature in their mass-radius relation.

The generic CSS EOS is not a very commonly used EOS for quark matter and is not suitable for applications in CCSN simulations, as it does not provide a temperature dependence or information about the composition. However, we show that it is possible to transform the CSS parameters into parameters of the widely used thermodynamic bag model, which does not have these deficits.

From the transformation of the CSS to the bag model EOS we identify that for certain quark matter parameters the problem of reconfinement can occur, where after a first deconfinement a spurious reconfinement and another deconfinement phase transition happen. 
We find that some other parameter regions actually correspond to absolutely stable strange quark matter.
The problem of reconfinement is known in the literature (e.g.\ \cite{2005ApJ...629..969A,lastowiecki2012,2013AA...553A..22C}), but in the parameter scans of Alford \textit{et al}.\ and subsequent works it was not addressed. If one does not consider reconfinement, and by doing so ignores thermodynamic stability in a strict sense, this leads to extremely high neutron star masses of over 3 M$_\odot$ at low transition pressures. In this work, we show the effects of reconfinement on the maximum mass in our parameter scan and that such high masses cannot be obtained any more if reconfinement is taken into account. Furthermore, for the first time we give a systematic analysis for which conditions the problem of reconfinement occurs.

Hyperons represent an additional degree of freedom which can be considered in the hadronic EOS. Their appearance generally leads to a softening of the EOS and therefore to a lower maximum mass. Often it is hard to even meet the 2~M$_\odot$ constraint. This problem is known under the name ``hyperon puzzle''; see, e.g., \cite{lonardoni2015,chatterjee15}. However, several hyperonic neutron star EOSs exist which have sufficiently high maximum masses by including repulsive hyperon interactions. An alternative solution to this puzzle is a phase transition to quark matter at low densities, which takes place before the appearance of hyperons; see \cite{schulze11,zdunik13,chatterjee15}. Regarding SN EOSs, there is only one model (the EOS of Ref.~\cite{2014ApJS..214...22B} named BHB$\Lambda \phi$ EOS) which is directly compatible with the measurement of \cite{2013Sci...340..448A}. It represents an extension of HS(DD2) where lambda hyperons have been added. 
We use this EOS in the present work to investigate the impact of hyperons on our parameter scan and the problem of reconfinement. We find that the overall results do not change qualitatively, besides the general reduction of the maximum mass.

The paper is structured as follows: In Secs. \ref{sec_classification}, \ref{sec_modeling}, and \ref{sec_parameterscan}, we repeat some hybrid star theory, present the models used and give detailed information on the performed parameter scan with Alford's classification. 
In Sec.\ \ref{sec_quarkmodels}, we show how to transfer the CSS EOS into a thermodynamic bag EOS. 
In Sec.\ \ref{sec_restricitingparamterspace}, we repeat our parameter scan using an extended parameter space. We identify interesting configurations for future SN EOS, while also comparing them with the already existing ones.
Section \ref{sec_reconfinement} deals with reconfinement, where we identify regions with one, two and three phase transitions. Another parameter scan is presented which shows the consequences of reconfinement on the maximum mass.
In the last section, Sec.\ \ref{sec_hyperons}, the detailed analysis of Secs.\  \ref{sec_restricitingparamterspace} and \ref{sec_reconfinement} is repeated applying the BHB$\Lambda \phi$ EOS for the hadronic part which additionally considers hyperons. The results are discussed in detail and compared with the nonhyperon EOS HS(DD2).
In Sec.\ \ref{sec_summary} we summarize and draw conclusions. Throughout the paper we use units where $k_{B}=\hbar=c=1$. 

\section{Classification of Hybrid Stars}
\label{sec_classification}
\begin{figure}
 \includegraphics[width = 1. \columnwidth]{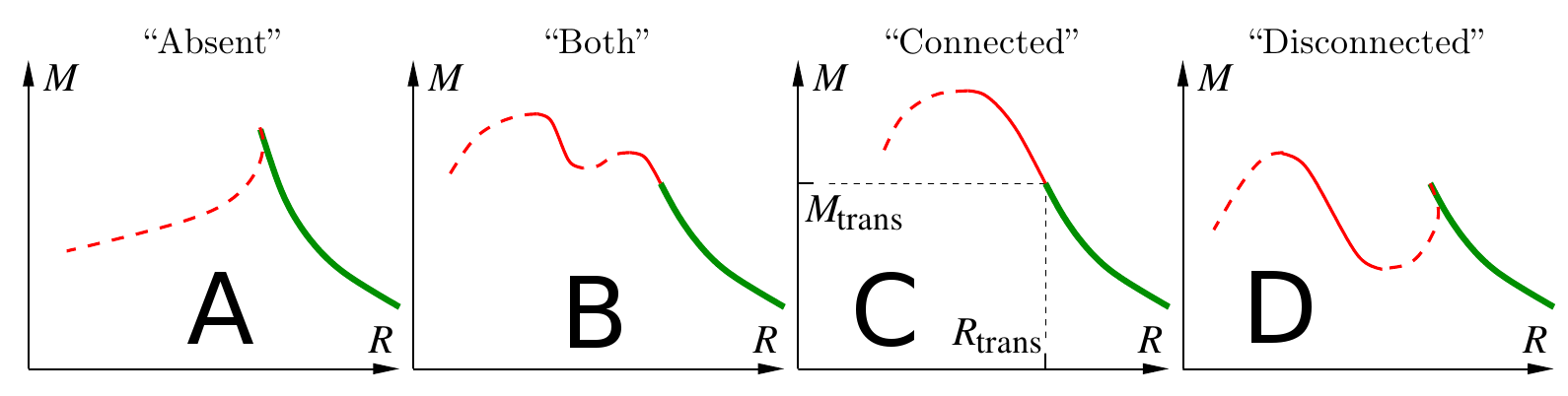}
 \caption{\label{fig_fourcases} This illustration was published in \cite{PhysRevD.88.083013} and shows the classification of hybrid stars by means of their $M$-$R$ curve. The two important criteria are the presence of a third family branch and the stability of hybrid stars at the onset of quark matter.}
\end{figure}

Alford \textit{et al}.\ introduced in Ref.~\cite{PhysRevD.88.083013} four different cases to classify hybrid stars by their $M$-$R$ relation as shown in Fig. \ref{fig_fourcases}. The classification is based on two criteria: the presence of a third family branch and the stability of hybrid stars at the onset of quark matter. Cases A and C have no third family branch and therefore only one maximum mass configuration.
Case A (``absent'') consists of only a hadronic branch. The point where quark matter sets in coincides with the maximum mass configuration. Case C (``connected'') is similar to case A with the difference that there are stable hybrid star configurations which include quark matter up to the maximum mass.
Cases B and D both have a third family branch in their $M$-$R$ curve. Case B is identical to case C up to the first maximum. There is an unstable branch to the left of this point, followed by a third family branch ending in a second maximum. Case D is identical to case A up to the first maximum, but also has a third family branch in addition.
For the supernova mechanism triggered by the hadron-quark phase transition, cases B and D are interesting. 
They both have the potential to induce a second collapse in a SN and a subsequent explosion as described in the Introduction.

\section{Hybrid Star Modeling}
\label{sec_modeling}
Alford \textit{et al}. introduced in Ref.~\cite{PhysRevD.88.083013} a simple model to describe hybrid stars in a systematic way. We closely follow this modeling except one difference: Alford \textit{et al}.\ used the rather soft HLPS and the rather stiff NL3 EOS in \cite{PhysRevD.88.083013} for the hadronic part (respectively BHF and DBHF in \cite{2016EPJA...52...62A}), to illustrate its impact on the hybrid star configurations. Instead we apply HS(DD2), which has a ``stiffness'' somewhere in between the EOSs used by Alford \text{et al}.
The quark phase is still described by the constant speed of sound (CSS) EOS as in Ref.~\cite{PhysRevD.88.083013}. Both phases are connected by the means of a Maxwell construction \cite{PhysRevD.88.083013}. In Sec.~\ref{sec_hyperons} we study the effect of hyperons by using the BHB$\Lambda\phi$ EOS. In the following a brief summary about the used EOSs is given.

\subsection{Hadronic matter: HS(DD2) and BHB$\Lambda\phi$}
The HS(DD2) EOS \cite{hempel10,Fischer2014} is a supernova EOS available at finite temperature and variable proton fraction and density in the form of a table. Nucleons and nuclei are considered as baryonic particle degrees of freedom. For the interactions of the nucleons density-dependent relativistic mean-field theory is used (DD2) \cite{PhysRevC.81.015803}. The formation of nuclei at subsaturation densities is considered in a statistical description employing experimentally measured binding energies and excluded-volume corrections \cite{hempel10}. 
The HS(DD2) EOS is in good agreement with experimental constraints for the symmetry energy \cite{hempel15b}, theoretical constraints for the neutron matter EOS \cite{Fischer2014}, and cluster formation in low-energy heavy-ion collisions \cite{hempel15}. Conversely, the EOS of Refs.~\cite{shen98,shen98_2,shen11} named STOS in the following, which is employed for the hadronic part in the currently existing quark-hadron hybrid SN EOSs (listed and further discussed in Sec.~\ref{sec_restricitingparamterspace}), is in contradiction with these constraints. HS(DD2) has a high maximum mass for cold neutron stars of 2.42~M$_\odot$.

The only existing SN EOS that considers hyperons and strictly fulfills the 2.01~M$_\odot$ neutron star constraint of \cite{2013Sci...340..448A} is the BHB$\Lambda \phi$ EOS \cite{2014ApJS..214...22B}. 
It represents an extension of HS(DD2) where the $\Lambda$ hyperon has been added as a particle degree of freedom within the density-dependent relativistic mean-field framework. Otherwise, the underlying models of HS(DD2) and BHB$\Lambda \phi$ are identical, e.g., regarding the nucleon interactions or the description of nuclei. 
Other hyperons than the $\Lambda$ are not considered in BHB$\Lambda \phi$. The justification of this simplification is that the experimental data for the interactions of the other hyperons are even more uncertain than they are for the $\Lambda$, and that often the $\Lambda$ is found to be the most important hyperon regarding the neutron star EOS. To reach the 2~M$_\odot$ constraint repulsive hyperon-hyperon interactions have been included in the BHB$\Lambda \phi$ EOS via the strange $\phi$ meson. The resulting maximum mass for cold, $\beta$-equilibrated matter is $2.11$~M$_\odot$, and thus directly compatible with the measurement of \cite{2013Sci...340..448A}. This means BHB$\Lambda \phi$ does not show a hyperon puzzle.

In this work, we use the HS(DD2) and BHB$\Lambda \phi$ EOSs in beta equilibrium and at $T=0.1$~MeV. 
A temperature of 0.1~MeV is negligibly small in comparison to typical Fermi energies in neutron stars, and thus a sufficient approximation for $T=0$. 
Note that the inner and outer crust is included self-consistently in HS(DD2) and BHB$\Lambda\phi$, i.e., we have a unified EOS description for the entire neutron star. 

\subsection{Quark matter: CSS}
The quark phase is described by the CSS EOS of Alford \textit{et al}.~\cite{PhysRevD.88.083013}:
\begin{equation}
 \epsilon_{\rm CSS}(p) = c_{\rm QM}^{-2}(p-p_{0}) \; ,
 \label{eq_CSS}
\end{equation}
where $c_{QM}$ is the density-independent speed of sound, $p$ the pressure, and $p_0$ the pressure where $\epsilon_{\rm CSS}=0$. Two values for $c_{\rm QM}$ are of special interest: $c_{\rm QM}^2=1/3$ which corresponds to non- or weakly interacting, massless quarks and $c_{\rm QM}^2=1$ which is the maximum value to be still consistent with special relativity. In this paper, $c_{\rm QM}^2=1/3$ is used, which is typical for many quark EOSs and also in agreement with other, more sophisticated models (e.g., \cite{2015PhRvD..92h3002A,2015AA...577A..40B}). 

\subsection{Hybrid EOS}
The phase transition from the hadronic phase described by the HS(DD2) or BHB$\Lambda \phi$ EOS to the quark phase described by the CSS EOS is done by a Maxwell construction. This means that local charge neutrality is assumed implicitly. It implies pressure, temperature and baryon chemical equilibrium at the transition point and that no phase coexistence region is present in compact stars. 
In fact, previous parameter scans did not consider chemical equilibrium explicitly, which we will discuss further in Sec.~\ref{sec_reconfinement}.
Pressure equilibrium at the transition pressure $p_{\rm trans}$ can be formulated as $p^{\rm hadronic}=p^{\rm quark}=p_{\rm trans}$.
A direct consequence of the Maxwell construction is the appearance of a discontinuity in the energy density $\Delta \epsilon = \epsilon^{\rm quark} - \epsilon^{\rm hadronic}$ at $p_{\rm trans}$. For a deconfinement transition from hadronic to quark matter one has $n_B^{\rm quark}>n_B^{\rm hadronic}$ (with the baryon number density $n_B$) and therefore also $\epsilon^{\rm quark}>\epsilon^{\rm hadronic}$.

The phase transition and the quark EOS depend on three variables: the transition pressure $p_{\rm trans}$, 
the speed of sound in quark matter $c_{\rm QM}$ and the value of the discontinuity in the energy density $\Delta \epsilon$. In the present work $c_{\rm QM}^2=1/3$ is fixed and $p_{\rm trans}$ and $\Delta \epsilon$ are varied systematically.
The final form of the EOS is written as
\begin{equation}
  \epsilon(p)=
  \begin{cases}
    \epsilon^{\rm hadronic}(p)&p \leq p_{\rm trans}\\
    \epsilon^{\rm hadronic}(p_{\rm trans}) + \\
             \Delta\epsilon + c_{\rm QM}^{-2}(p-p_{\rm trans})&p>p_{\rm trans}  \; .
  \end{cases}
  \label{hybrideos}
\end{equation}
This means that $p_0$ in Eq.~(\ref{eq_CSS}) is fixed by the pressure and energy density of quark matter at the transition point, $p_{\rm trans}$ and $\epsilon_{\rm CSS}(p_{\rm trans})=\epsilon_{\rm trans} + \Delta\epsilon$, with $\epsilon_{\rm trans}=\epsilon^{\rm hadronic}(p_{\rm trans})$, 
leading to
\begin{equation}
\label{eq_p0}
 p_0 = p_{\rm trans} - c_{\rm QM}^2(\epsilon_{\rm trans} + \Delta \epsilon).
\end{equation}

\section{Parameter Scan}
\label{sec_parameterscan}
Two pieces of information are especially relevant when modeling a hybrid star: its maximum mass and the type of hybrid star.
To calculate a single compact star, the Tolman-Oppenheimer-Volkoff equations have to be solved for a given central density:
\begin{eqnarray}
 \frac{dp}{dr} &=& - \frac{G \epsilon(r) m(r)}{r^2} \left( 1+\frac{p(r)}{\epsilon(r)} \right) \nonumber \\
 && \times\left( 1+\frac{4\pi r^3 p(r)}{m(r)} \right) \left(1-\frac{2Gm(r)}{r} \right)^{-1} \; , \\
 \label{tov}
 \frac{dm}{dr} &=& 4 \pi r^2 \epsilon(r) \quad ,
 \label{3secondeqstr}
\end{eqnarray}
with the enclosed mass $m$ at radius $r$ and the gravitational constant $G$.
The maximum mass configuration with fixed $p_{\rm trans}$ and $\Delta \epsilon$ is obtained from the $M$-$R$ relation, where the central density of the hybrid stars is systematically varied. If $p_{\rm trans}$ and $\Delta \epsilon$ are systematically varied as well, a three-dimensional surface plot of the maximum mass as a function of these two parameters is obtained. 80 variations of each $p_{\rm trans}$ and $\Delta \epsilon$ are considered here, 
varying $p_{\rm trans}$ from 1~MeV$/$fm$^{3}$ ($n_B \approx 0.1$ fm$^{-3}$) to 800~MeV$/$fm$^{3}$ ($n_B \approx 1.02$~fm$^{-3}$) while using HS(DD2) EOS. $p_{\rm trans}$ also fixes $\epsilon_{\rm trans}$, resulting in values $p_{\rm trans}/\epsilon_{\rm trans}=[0.01,0.55]$. 
$\Delta \epsilon / \epsilon_{\rm trans}$ is varied within the range $[0,1.3]$. In Sec.\ \ref{sec_restricitingparamterspace} we will also present an extended parameter scan for HS(DD2), covering the range of $p_{\rm trans}/\epsilon_{\rm trans}=[0.01,0.55]$ and $\Delta \epsilon / \epsilon_{\rm trans} = [0,3]$. For BHB$\Lambda \phi$, $p_{\rm trans}$ is varied from 1 MeV/fm$^3$ to 640 MeV/fm$^3$, while $\Delta \epsilon / \epsilon_{\rm trans}$ is varied from $[0,1.3]$.

\begin{figure}
 \includegraphics[width = 1. \columnwidth]{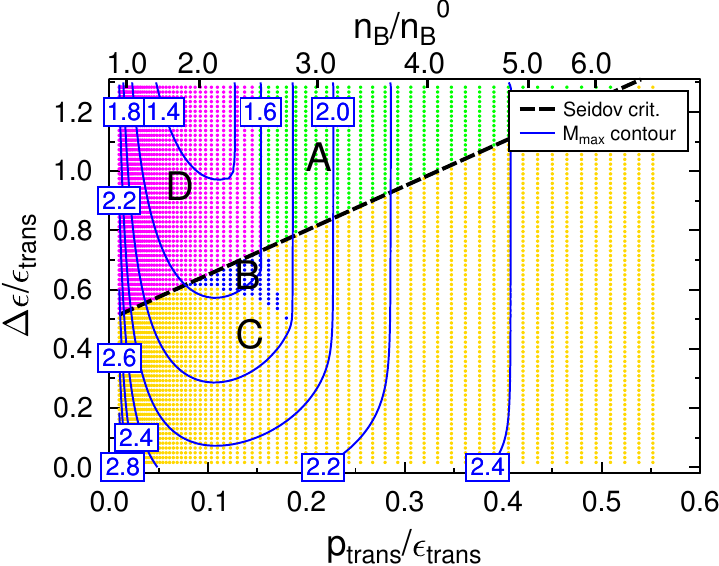}
 \caption{\label{4Alfordcases} Calculated hybrid star configurations, colored to distinguish the four cases A (absent), B (both), C (connected) and D  (disconnected). The lines in blue show the maximum mass contours for 1.4, 1.6, 1.8, 2.0, 2.2, 2.4, 2.6, and 2.8 M$_\odot$. The thick black dashed line shows the analytic criterion from Seidov \cite{1971SvA....15..347S}, above which neutron stars are unstable at the onset of quark matter.}
\end{figure}

Figure \ref{4Alfordcases} shows contour lines of the maximum mass for our considered range of parameters.
The most important contour line is the 2 M$_\odot$ mass line, since all EOSs have to be able to support this mass.
Such heavy compact stars can be reached at $p_{\rm trans}/\epsilon_{\rm trans} \gtrsim 0.22$ (case 1) and $p_{\rm trans}/\epsilon_{\rm trans} \lesssim 0.02$ (case 2) for any $\Delta \epsilon / \epsilon_{\rm trans}$. For $ 0.02 \lesssim p_{\rm trans}/\epsilon_{\rm trans} \lesssim 0.22$, $\Delta \epsilon / \epsilon_{\rm trans}$ (case 3) is limited to low values to be compatible with the observational constraint. 
In case 1, the hadronic phase is dominant and a mass of 2 M$_\odot$ is reached already in the hadronic branch. The higher $p_{\rm trans}/\epsilon_{\rm trans}$ gets, the later the quark phase sets in. At high values of $p_{\rm trans}/\epsilon_{\rm trans}$, hybrid stars consist almost only of hadronic matter. For $p_{\rm trans}/\epsilon_{\rm trans}>0.47$, eventually the transition pressure is above the central pressure of the heaviest stable hadronic star. For low $p_{\rm trans}/\epsilon_{\rm trans}$ (case 2), one obtains an almost pure quark star with only a thin hadronic layer on top.
At the lowest $p_{\rm trans}$ and $\Delta \epsilon$, extremely high maximum masses of over 3 M$_\odot$ can be reached,
well above the maximum mass of HS(DD2).

The dots in Fig.~\ref{4Alfordcases} represent all the parameter configurations that
have been calculated. The color coding classifies the resulting $M$-$R$ relations according to the four cases of Alford \textit{et al}. 
The straight black diagonal line represents the analytical constraint derived by Seidov in 1971 \cite{1971SvA....15..347S}: $\Delta \epsilon_{\rm crit}/\epsilon_{\rm trans} = 1/2+3/2 \cdot p_{\rm trans}/\epsilon_{\rm trans}$. If $\Delta \epsilon$ is below $\Delta \epsilon_{\rm crit}$, hybrid stars are stable at the onset of quark matter.
Above the Seidov line, cases A (green) and D (magenta) are found, below cases C (yellow) and B (blue). Apparently, $p_{\rm trans}$ has to be chosen low enough, to obtain a disconnected third family branch. 
Interesting cases for SN simulations are in the small region on the left side of the two solar mass line and above the Seidov line. There, hybrid stars with a third family branch and maximum masses above 2~M$_\odot$ are found. Furthermore, they correspond to low onset densities of the phase transition between 1 and 2 $n_B^0$ (with $n_B^0$ denoting the nuclear saturation density) which is required to reach quark matter in a CCSN, at least for low- and intermediate-mass progenitors, see \cite{fischer12}. Note that such low transition densities are compatible with heavy-ion collision experiments, where matter is more symmetric and strangeness is not in equilibrium, which shifts the phase transition to much higher densities \cite{2011ApJS..194...39F}.

In \cite{2016EPJA...52...62A}, Alford and Han showed results of a similar parameter scan done also for the CSS EOS with $c_s^2 = 1/3$, but with the stiff DBHF EOS and the soft BHF EOS for the hadronic phase, and in \cite{PhysRevD.88.083013} for the HLPS and NL3 hadronic EOSs. 
The general distribution of the cases found here is the same as in \cite{PhysRevD.88.083013} and \cite{2016EPJA...52...62A}. 
The 2~M$_\odot$ curve from Fig.~\ref{4Alfordcases} behaves in a similar way as the 1.95~M$_\odot$ line of DBHF in Fig.~5 of \cite{2016EPJA...52...62A}. Considering these two references, our results seem to be consistent with Alford \textit{et al}. We can also state here, that for $c_{\rm QM}^2 = 1/3$, the hadronic phase has little impact on the distribution of the hybrid stars in the $p_{\rm trans} / \epsilon_{\rm trans}$ vs $\Delta \epsilon / \epsilon_{\rm trans}$ plot.

\section{Quark EOS Models}
\label{sec_quarkmodels}
The CSS EOS is not a very common EOS for the description of quark matter. Furthermore, as it only represents a parametrization of thermodynamic quantities, it does not contain any composition or temperature dependence. Both aspects are important for the application in CCSN simulations which we are aiming at.
A commonly used and easy-to-handle model which provides this information is the so-called thermodynamic bag model, which is described in more detail in the following.
In 1984, Witten proposed the concept of absolutely stable strange quark matter \cite{PhysRevD.30.272}. In the same year, Farhi and Jaffe investigated Witten's theory by using a Fermi-gas model to establish conditions under which strange matter in bulk is absolutely stable. They considered three-flavor ($u$, $d$, $s$) quark matter in beta equilibrium at zero temperature with a negative external bag pressure $B$ acting on quark matter \cite{1984PhRvD..30.2379F}. Matter is assumed to be in equilibrium regarding the following reactions:
\begin{eqnarray}
d \leftrightarrow u+e+\bar{\nu_e} \nonumber \; ,\\
s \leftrightarrow u+e+\bar{\nu_e} \nonumber \; ,\\
s + u \leftrightarrow u + d \; .
\end{eqnarray}
In cold neutron stars where no neutrinos are present the chemical potentials thus fulfill the relation:
\begin{equation}
\label{eq_murelations}
 \mu_d=\mu_s = \mu_u + \mu_e \; .
\end{equation}

The pressure $p_i$ depending on the chemical potential $\mu_i$ for each species $i= u,d,s,e$ is easily calculated since they are treated as noninteracting Fermi gases:
\begin{eqnarray}
 p_i &=& \frac{1}{6}\frac{g}{4\pi^2} \left[\mu_i(\mu_i^2-m_i^2)^{1/2}(\mu_i^2-\frac{5}{2}m_i^2) \right. \nonumber\\
 && \left. +\frac{3}{2}m_i^4\ln\left(\frac{\mu_i^2-m_i^2)^{1/2}+\mu_i}{m_i}\right)\right] \; .
\end{eqnarray}
The degeneracy factor $g$ is $g = 2_{\rm spin}$ for electrons and $g = 6 = 2_{\rm spin} \times 3_{\rm color}$ for quarks. The pressures for each species, assuming the masses for $u$ and $d$ quarks as well as electrons are negligible, are:
\begin{eqnarray}
\label{eq_farhijaffe_noninteracting}
 p_u^{\rm non-int} &=& \frac{\mu_u^4}{4\pi^2} \nonumber \; ,\\
 p_d^{\rm non-int} &=& \frac{\mu_d^4}{4\pi^2} \nonumber \; ,\\
 p_e^{\rm non-int} &=& \frac{\mu_e^4}{12\pi^2} \nonumber \; ,\\
 p_s^{\rm non-int} &=& \frac{1}{4\pi^2}\left[\mu_s(\mu_s^2-m_s^2)^{1/2} (\mu_s^2-\frac{5}{2}m_s^2)\right. \nonumber\\
 && \left. +\frac{3}{2}m_s^4 \ln\left(\frac{(\mu_s^2-m_s^2)^{1/2}+\mu_s}{m_s}\right)\right] \; .
\end{eqnarray}
The total pressure is the sum of the particle pressures with the bag constant subtracted:
\begin{equation}
 p_{\rm tot} =  \sum_i p_i^{\rm non-int} - B \; .
 \label{eq_ptot}
\end{equation}
By using the number density for each species $n_i$, which can be obtained from the thermodynamic relation
\begin{equation}
\label{eq_ni}
 n_i = \frac{\partial p_{\rm tot}}{\partial \mu_i}\; , 
\end{equation}
the charge neutrality condition can be expressed as
\begin{equation}
\label{eq_chargeneutrality}
 \frac{2}{3}n_u - \frac{1}{3}n_d - \frac{1}{3}n_s - n_e = 0 \; .
\end{equation}

Equations (\ref{eq_chargeneutrality}) and (\ref{eq_murelations}) leave only one independent chemical potential. Using the $T=0$ thermodynamic relation
\begin{equation}
 \epsilon_{\rm tot} = -p_{\rm tot} + \sum_i  \mu_i n_i
 \label{eq_epsrelation}
\end{equation}
and  Eq.~(\ref{eq_ptot}), the total energy density can be written as
\begin{eqnarray}
 \epsilon_{\rm tot} &=& \sum_i(-p_i^{\rm non-int} + \mu_i n_i) + B \\
 &=& \sum_i\epsilon_i^{\rm non-int}  + B \; . 
\end{eqnarray}

To include interactions, often a phenomenological parametrization is used. Here, we apply the model of \cite{2011ApJS..194...39F} for $T=0$:
\begin{equation}
  p^{\rm QM}= \sum_i p_i^{\rm non-int}-B - \sum_{j=u,d,s} \frac{2 \alpha_s}{\pi}\frac{\mu_{j}^4}{4 \pi^2} , \;
  \label{eq_pqm}
\end{equation}
where $\alpha_s$ accounts for strong interaction corrections.
The model presented in \cite{2011ApJS..194...39F} is similar to the ones from Alford \textit{et al}.\ \cite{2005ApJ...629..969A} and Weissenborn \text{et al}.\ \cite{2011ApJ...740L..14W}. Both use an interaction correction proportional to $\mu^4$ (where $\mu$ denotes the quark chemical potential) similar to the $\alpha_s$ term in Eq.~(\ref{eq_pqm}). In fact, Weissenborn's model is equivalent to Eq.~(\ref{eq_pqm}) for $m_s=0$, and in this case the proportionality factor $a_4$ of the $\mu^4$-term can be identified as
 $a_4 = 1 - 2\alpha_s/\pi$. 
Alford's quark EOS is a generic power-series ansatz, which includes an additional $a_2\mu^2$ term.
This term can be interpreted to be related to color superconductivity by using the relation $a_2 = m_s^2-4\Delta^2$, where $\Delta$ represents the pairing gap \cite{alford03,2005ApJ...629..969A}.  
Another quark model suitable for the astrophysical application is vBag, which was introduced in \cite{klaehn15,klaehn16}. It contains vector interactions and a medium-dependent bag pressure, which is based on the assumption of simultaneous deconfinement and chiral symmetry restoration. It would be interesting to compare vBag with the quark EOSs used in the present study in the future.

An important case is where $u$, $d$, and $s$ quarks are massless. 
It follows $\mu_u = \mu_d = \mu_s = \mu$, and $\mu_e=0$, and $n_u = n_d = n_s$ and $n_e = 0$, i.e., quarks maintain charge neutrality by themselves and there are no electrons in the quark phase.
To be able to compare the CSS EOS with the bag model of Eq.~(\ref{eq_pqm}) in the limit of $m_s=0$ and beta equilibrium, Eq.~(\ref{eq_CSS}) has to be reformulated. Together with Eq.~(\ref{eq_epsrelation}), Eq.~(\ref{eq_CSS}) leads to
\begin{equation}
\label{eq_CIG1}
 p^{\rm CSS} = \frac{c_{\rm QM}^2}{1+c_{\rm QM}^2}\left(\frac{p_0}{c_{\rm QM}^2}  + \mu n\right) \; ,
\end{equation}
where $n=n_u+n_d+n_s$. $n$ depends on $\mu$ due to the relation $n = \partial p / \partial \mu$, which can be implemented in Eq.~(\ref{eq_CIG1}). Separating the variables and integrating over the respective boundaries leads to
\begin{equation}
 p^{\rm CSS}(\mu) = \frac{c_{\rm QM}^2}{1+c_{\rm QM}^2} p_0 \left[\left(\frac{\mu}{\mu_0}\right)^{\frac{1+c_{\rm QM}^2}{c_{\rm QM}^2}} + \frac{1}{c_{\rm QM}^2} \right] \; .
 \label{eq_pcss}
\end{equation}
It is interesting to note that another constant $\mu_0$ appears. 
The reason is that the $\epsilon(p)$-relation of Eq.~(\ref{eq_CSS}) does not represent a thermodynamic potential. For given $\Delta \epsilon$ and $p_{\rm trans}$, which fix $p_0$ by Eq.~(\ref{eq_p0}), $\mu_0$ can be fixed as well by inverting Eq.~(\ref{eq_pcss}) and using the condition of chemical equilibrium at the phase transition point,
\begin{equation}
 \mu^{\rm CSS}(p_{\rm trans})=\frac13 \mu_B^{\rm hadronic}(p_{\rm trans}) \; ,
 \label{eq_mueq}
\end{equation}
which gives
\begin{equation}
 \mu_0 = \frac{1}{3}\mu_B^{\rm hadronic}(p_{\rm trans})\left(\frac{1+c_{\rm QM}^2}{c_{\rm QM}^2}\frac{p_{\rm trans}}{p_0}-\frac{1}{c_{\rm QM}^2}\right)^{-\frac{c_{\rm QM}^2}{1+c_{\rm QM}^2}} \; .
\end{equation} 
The schematic form of Eq.~(\ref{eq_pcss}) was already given in the appendix of \cite{PhysRevD.88.083013}. However, in \cite{PhysRevD.88.083013} chemical equilibrium was not considered explicitly, as it is done above. This will be important in Sec.~\ref{sec_reconfinement}.
 
\begin{figure}
  \includegraphics[width = 1. \columnwidth]{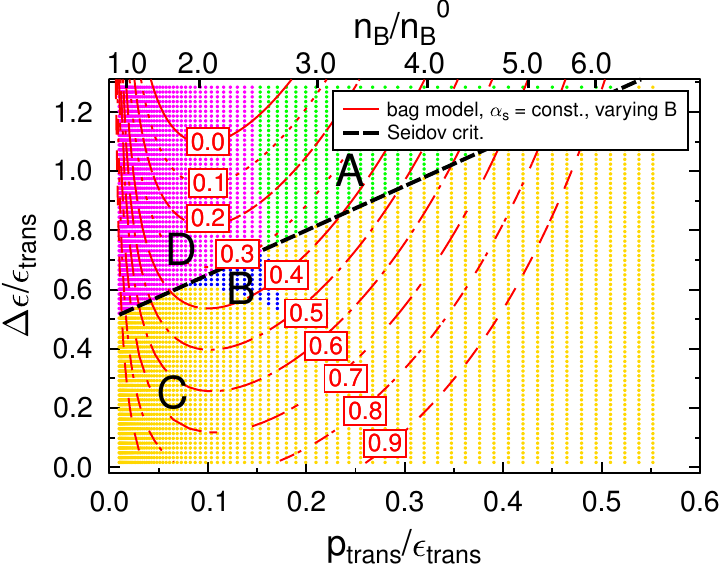}
  \caption{\label{cig_bag_identification} Similar to Fig.~\ref{4Alfordcases}, but with red lines showing the solutions of the bag model from Eq.~(\ref{eq_pqm}) for varying $B$ with increasing values from left to right, different values of $\alpha_s$ (as indicated in the figure), and $m_s=0$.}
\end{figure}

Comparing the bag model description of Eq.~(\ref{eq_pqm}) with the $p(\mu)$ formulation of the CSS EOS [Eq.~(\ref{eq_pcss})], it is obvious that these two formulations are equivalent when $c_{\rm QM}^2=1/3$.
The identifications of the $\mu^4$-dependent and $\mu$-independent terms in the CSS and bag EOSs lead to 
\begin{eqnarray}
 \alpha_s &=& \frac{\pi}{2} -\frac{\pi^3}{6}\frac{p_0}{\mu_0^4} \nonumber \\
  B &=&  - \frac{3}{4} p_0 \; .
\end{eqnarray}
Figure \ref{cig_bag_identification} shows a comparison of the two models. By varying the bag constant $B$ from lower to higher values (left to right on the red curves) as well as the $\alpha_s$ parameter (increasing $\alpha_s$ leads to a downward shift of the curves), the whole parameter space of the CSS model can be reproduced. 

\begin{figure}[h]
 \centering
 \includegraphics[width = \columnwidth]{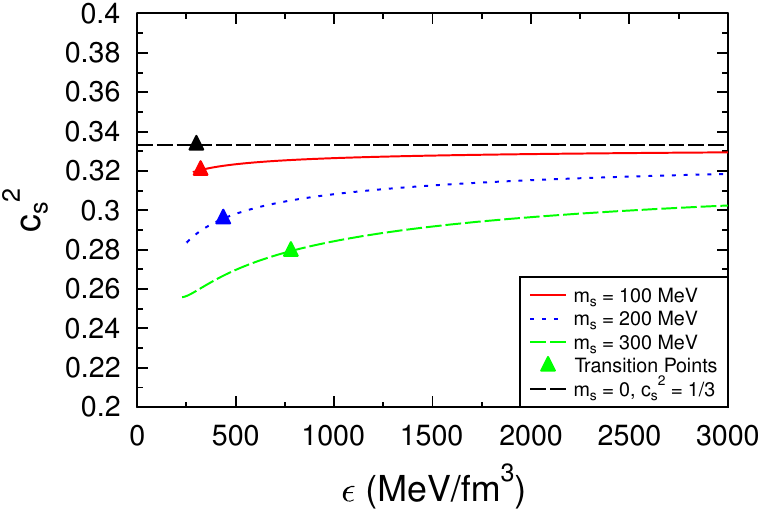}
 \caption{Dependency of the speed of sound on the energy density for four different values of $m_s$ (0, 100, 200 and 300 MeV), $B^{1/4} = 155 $~MeV and $\alpha_s = 0.3$. The phase-transition points are indicated with triangles.}
 \label{fig_cs_vs_mass}
\end{figure}

More realistic models of quark matter often employ a finite strange quark mass. A typical value is $m_s = 100$ MeV, which is, as mentioned by Fischer \textit{et al}.\ in \cite{2011ApJS..194...39F} in accordance with the range $m_s \sim 70 - 130$ MeV and the weighted average of $105^{+1.5}_{-1.3}$ MeV of Amsler \text{et al}.\ \cite{2008PhLB..667....1A}. 
Figure \ref{fig_cs_vs_mass} shows the influence of a finite $m_s$ on the speed of sound squared $c_s^2$. With increasing $m_s$, the speed of sound deviates significantly from the value of $c_s^2=1/3$, corresponding to $m_s = 0$~MeV. However, for $m_s = 100$~MeV the deviations are still small. 
The energy densities $\epsilon_{\rm trans}+\Delta \epsilon$ at the phase transition from hadronic to quark matter are indicated in Fig.~\ref{fig_cs_vs_mass} by triangles. For $m_s = 200$ and 300 MeV, the strongly deviating part at the beginning is not of importance, since the phase transition happens at higher energy densities.
As visible in the figure, if the value of the strange quark mass is larger than 100~MeV, it shifts the phase transition to higher densities, but for $m_s=100$~MeV the effect is still small. 
As a conclusion we can state that with a finite $m_s$ the one-to-one correspondence between the CSS model and the bag model is not true anymore, but nevertheless the models are still comparable. We have checked that at least for $m_s = 100$~MeV the induced differences in the $M$-$R$ relation are small. Only for detailed comparisons, the exact $M$-$R$ relations have to be calculated with the strange quark mass taken into account.

\section{Restricting the bag model parameter space}
\label{sec_restricitingparamterspace}
Sagert \textit{et al}.\ and Fischer \textit{et al}.\  \cite{2011ApJS..194...39F,2009PhRvL.102h1101S,sagert12} generated several hybrid SN EOSs in their papers. Here we use the same quark interactions as applied in some of these EOSs.
Similar quark-hadron hybrid SN EOSs have also been generated by Nakazato \textit{et al}.\ \cite{2008PhRvD..77j3006N,2013AA...558A..50N}. They did not consider corrections from strong interactions, and therefore obtain maximum masses only below 2~M$_\odot$.
Table~\ref{tab_sagert} gives an overview of the already published hybrid SN EOSs. 
In all of them, STOS \cite{shen98,shen98_2,shen11} was used for the hadronic part.
\begin{table}
\begin{tabular}{c c c c c c l}
\hline 
Name & $B^{1/4}$ & $\alpha_s$  & $M_{\rm max}$ & Explosion & Reference  \\ 
     & (MeV)     &             & (M$_\odot$)   &           & \\
\hline 
  B162 & 162 & 0    & 1.56 & Yes & \cite{2009PhRvL.102h1101S,2011ApJS..194...39F} \\
  B165 & 165 & 0    & 1.50 & Yes & \cite{2009PhRvL.102h1101S,2011ApJS..194...39F}\\
  B155 & 155 & 0.3  & 1.67 & Yes & \cite{2011ApJS..194...39F}\\ 
  B139 & 139 & 0.7  & 2.04 & No  & \cite{sagert12,2014_fischer_polonica}  \\ 
  B145 & 145 & 0.7  & 1.97 & No  & \cite{sagert12} \\ 
  B209 & 209 & 0    & 1.80 & No  & \cite{2008PhRvD..77j3006N,2013AA...558A..50N} \\ 
  B162 & 162 & 0    & 1.54 & Yes & \cite{2013AA...558A..50N} \\
  B184 & 184 & 0    & 1.36 & No  & \cite{2013AA...558A..50N} \\
\hline
\end{tabular}
\caption{\label{tab_sagert} Overview of existing hybrid SN EOSs and their tests in spherically symmetric CCSN simulations. All models employ m$_s=100$~MeV.}
\end{table}

As summarized in Table~\ref{tab_sagert}, so far only hybrid EOSs that have maximum masses below 2 M$_\odot$ were found to lead to explosions in spherically symmetric CCSN simulations. In particular, the models B139 and B145, which both have QCD interaction terms and support maximum masses around 2 M$_\odot$, did not lead to explosions. Currently, these are the only two available SN EOSs that include quark matter and support maximum neutron stars masses above $2$ M$_\odot$. Note that so far only very few progenitors have been tested in CCSN simulations of this scenario. A systematic progenitor exploration is still missing, even for the few existing hybrid SN EOSs.

\begin{figure}
    \includegraphics[width = 1. \columnwidth]{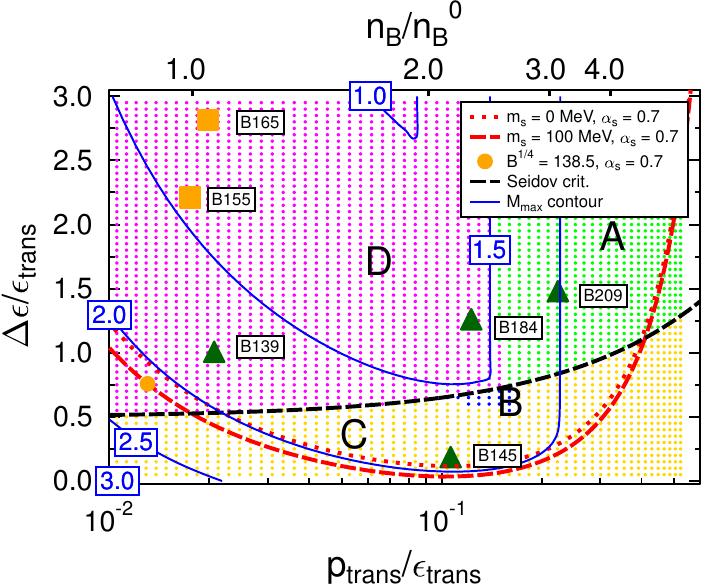}
\caption{\label{search5} Parameter scan extended to higher  $\Delta \epsilon/\epsilon_{\rm trans}$ and logarithmic scale for $p_{\rm trans}/\epsilon_{\rm trans}$. Additionally, the phase transition parameters of the hybrid EOSs of Table~\ref{tab_sagert} are plotted with yellow squares (explosions found) and green triangles (no explosions found). Note that the hadronic part of these EOSs is based on STOS, whereas the results in the figure (maximum mass and classification) are based on HS(DD2), for details see the main text. Marked with a yellow circle is an example case whose $M$-$R$ relation is shown in Fig.~\ref{fig_ms100a4148}.
} 
\end{figure} 

In the following, we use the properties of the existing hybrid SN EOSs listed in Table~\ref{tab_sagert} to identify interesting regions of the quark matter parameter space which could be favorable for CCSN explosions.
In these EOSs STOS is used for the description of hadronic matter and Gibbs' conditions for phase equilibrium are applied. For the present parameter scan HS(DD2) and Maxwell's conditions are used instead, which complicates the comparison. For the aspects we are mostly interested in the parameters $p_{\rm trans}/\epsilon_{\rm trans}$ and $\Delta \epsilon/\epsilon_{\rm trans}$ are more relevant than the bag model parameters: the former have a physical meaning independent on the particular hadronic EOS that is used, as, for example, they determine whether or not hybrid stars are stable at the onset of quark matter (cases A and D vs B and C). Therefore we calculate these parameters for the STOS EOS and the given bag model parameters $m_s$, $B$ and $\alpha_s$. As the only difference to the original hybrid EOSs of Table~\ref{tab_sagert}, we have to assume local instead of global charge neutrality to achieve the desired Maxwell transition at constant pressure.
The results are shown by green triangles and yellow squares in Fig.~\ref{search5}.
Four of the seven configurations did not lie in the original parameter space used in Fig.~\ref{4Alfordcases}. For this reason, we expanded the parameter space to $\Delta \epsilon / \epsilon_{\rm trans}$ up to 3. Now only B162, with $\Delta \epsilon / \epsilon_{\rm trans} \approx 6.1$ and $p_{\rm trans}/\epsilon_{\rm trans}\approx 0.005$ lies outside the parameter range considered in the figure.
We are also using a logarithmic scale for $p_{\rm trans}/\epsilon_{\rm trans}$ to achieve a clearer presentation of the data. Note that the maximum mass contour lines and Alford classification are still calculated for the HS(DD2) EOS (as before) so that they are different from the values given in Table~\ref{tab_sagert}, due to the procedure described above.

The exploding EOSs B155 and B165 have values just slightly above $p_{\rm trans} / \epsilon_{\rm trans} = 0.01$. 
B139 has a comparable value of $p_{\rm trans} / \epsilon_{\rm trans}$ to the ones from B155 and B165, but differs in the energy discontinuity $\Delta \epsilon$ which is smaller. 
B145 seems to be rather different: the phase transition happens at very high $p_{\rm trans}/\epsilon_{\rm trans}$ and low $\Delta \epsilon /\epsilon_{\rm trans}$. B184 and B209 have similar $p_{\rm trans}/\epsilon_{\rm trans}$ but higher $\Delta \epsilon / \epsilon_{\rm trans}$.
These results indicate that a high $\Delta \epsilon / \epsilon_{\rm trans}$ and low $p_{\rm trans}/\epsilon_{\rm trans}$ are more favorable for obtaining explosions. Interestingly, these are the conditions that result in a disconnected third family. This confirms our expectations presented in the Introduction and is in agreement with \cite{2015arXiv151106551H}, that SN explosions induced by a QCD phase transition are related to the existence of a third family. Note that B139 has a disconnected third family but did not explode, indicating that a \textit{pronounced} third family is favorable for explosions. It has to be emphasized that the inclusion of the existing hybrid EOS in Fig.~\ref{search5} can only serve as a weak guideline regarding the explodability, 
because in the simulations a different hadronic EOS STOS is used, and global instead of local charge neutrality is assumed.

As already discussed in Sec.~\ref{sec_parameterscan}, the 2 M$_\odot$ line in Fig.~\ref{search5} excludes a lot of potential parameter combinations for new SN EOSs. 
Only the ``disconnected'' cases D in the lower left corner, which have a sufficiently high maximum mass, are left as interesting candidates. The other parameter regions with $M_{\rm max}>2$~M$_\odot$ have either a very low $\Delta \epsilon / \epsilon_{\rm trans}$ or a very high $p_{\rm trans} / \epsilon_{\rm trans}$, and in any case do not lead to a third family of compact stars. These results nicely illustrate the tension between high maximum masses and the possibility of CCSN explosions induced by a strong phase transition, but there is still an interesting parameter region remaining. 

Next we discuss the implications for the bag model formulation of the quark EOS. Choosing $\alpha_s = 0.7$ leads to configurations that lie almost on top of the 2~M$_\odot$ line in the lower left corner, as can be seen by the red dotted line in Fig.~\ref{search5}. We consider this as a lower boundary for $\alpha_s$ to choose. Higher values of $\alpha_s$ are allowed, too, but are constrained to be above the Seidov line if one requires a third family.
Considering a finite strange quark mass of $m_s = 100$~MeV shifts the $\alpha_s = 0.7$ line slightly to lower $\Delta \epsilon_{\rm trans}$ values, as can be seen by comparing with the red dashed line. However, cases with same bag constants $B$ and interaction parameters $\alpha_s$, but different strange quark masses $m_s$, can lead to big differences in $\Delta \epsilon / \epsilon_{\rm trans}$ and $\epsilon_{\rm trans}/p_{\rm trans}$ values, which is not visible in the figure.

\begin{figure}
    \includegraphics[width = 1. \columnwidth]{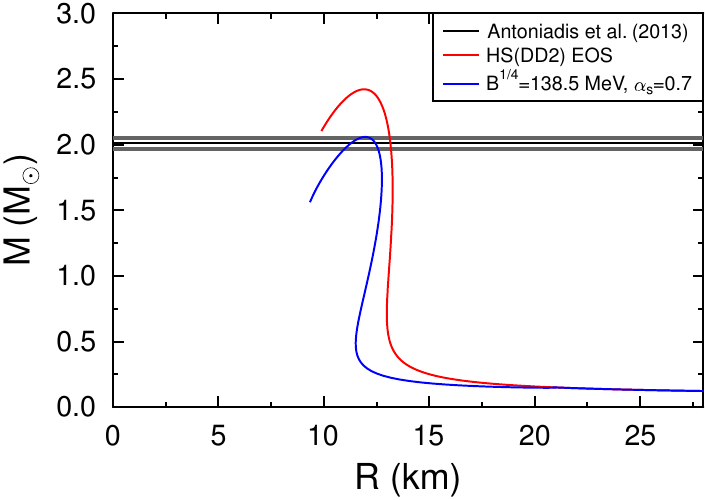}
\caption{\label{fig_ms100a4148} $M$-$R$ relation of an example hybrid EOS which might be interesting for CCSNe. The quark matter parameters are $m_s = 100$~MeV, $\alpha_s = 0.7$, and $B^{1/4}=138.5$~MeV.}
\end{figure}

As an example of what a possible hybrid star configuration might look like, we chose the configuration $m_s = 100$~MeV, $\alpha_s = 0.7$ and $B^{1/4}=138.5$~MeV. The phase transition properties are shown in Fig.~\ref{search5} by the yellow circle and the mass-radius relation is shown in Fig.~\ref{fig_ms100a4148}. The values of the phase transition parameters are $p_{\rm trans} / \epsilon_{\rm trans}=0.013$ and $\Delta \epsilon / \epsilon_{\rm trans}$=0.76. The maximum mass configuration has $M_{\rm max} = 2.05$~M$_\odot$ with $R = 11.98$~km. For 1.4~M$_\odot$ the hybrid EOS leads to a somewhat smaller radius of 12.64~km than HS(DD2) with 13.22~km. 
The onset of quark matter in the $M$-$R$ curve takes place around 22~km, corresponding to a density of 0.127~fm$^{-3}$. Note again, that such a low onset density in neutron stars is not in disagreement with heavy-ion collision experiments. For the conditions in heavy-ion collisions (isospin symmetric matter with zero net strangeness), the onset density 
at $T=0$ shifts to much higher values: for the example case to 0.962~fm$^{-3}$.

It is important to point out that the third-family feature of the example case is so weak, that it is almost not visible in Fig.~\ref{fig_ms100a4148}. Also for the other cases B and C with stable hybrid stars we found that the characterizing features often are very weak, and look very different than the prime examples of Fig.~\ref{fig_fourcases}. However, in \cite{2015arXiv151106551H} it was shown that finite entropies as they occur in the protoneutron star in a CCSN can significantly enhance the third-family features so that they become very pronounced. 

\section{Reconfinement and stability of quark matter}
\label{sec_reconfinement}
\begin{figure}
    \includegraphics[width = 1. \columnwidth]{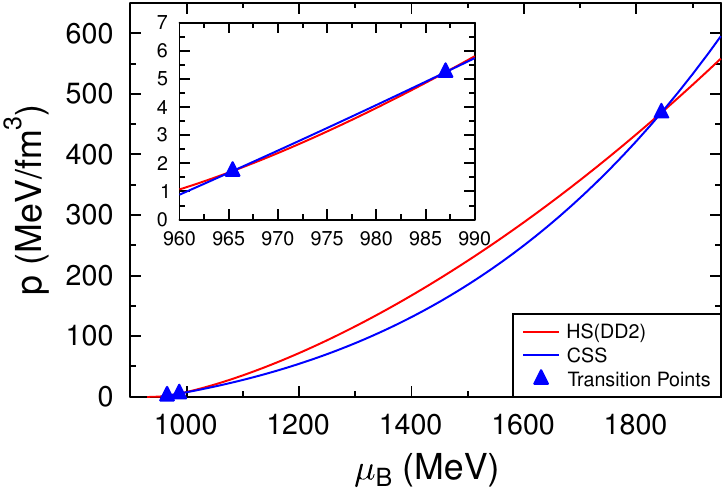}
\caption{\label{fig_p_mu_3PT_fullrange} Example case with three phase transitions. The inlay shows a zoom-in of the first two phase transitions.}
\end{figure}

In the model used in Sec.~\ref{sec_parameterscan}, by construction there is always just one (deconfinement) phase transition, which goes from hadronic to quark matter. Figure \ref{4Alfordcases} shows that in this case masses well above 2 M$_\odot$ and even above the maximum mass of the hadronic HS(DD2) EOS are possible. 
However, the $p(\mu_B)$ relation of the CSS EOS derived in Eq.~(\ref{eq_pcss}) reveals that more than a single phase transition can happen. 
Multiple phase transitions were also found for other hybrid EOSs; see, e.g., \cite{lastowiecki2012,zdunik13,2013AA...553A..22C,blaschke16}.
Figure \ref{fig_p_mu_3PT_fullrange} shows an example where three phase transitions occur. The phase transition in the original setup of the parameter scan, where the $p(\mu_B)$-relation is not considered, is the one most to the left, with values $p_{\rm trans}/\epsilon_{\rm trans} \simeq 0.014$ and $\Delta \epsilon/\epsilon_{\rm trans}=0.2$. For higher chemical potentials, by construction the CSS quark EOS is always used. From the selected values of $p_{\rm trans}$ and $\Delta \epsilon$ and the condition for chemical equilibrium at the transition point [Eq.~(\ref{eq_mueq})], the $p(\mu_B)$ relation of the CSS EOS is uniquely fixed. 
By using this relation as shown in Fig.~\ref{fig_p_mu_3PT_fullrange}, it turns out that quark matter is not the true ground state for chemical potentials between approximately 987 and 1844~MeV. Instead, at 987~MeV a reconfinement transition from quark to hadronic matter takes place, and another deconfinement transition around 1844~MeV. We abbreviate such a series of phase transitions as HQHQ. 
The original setup is forced to have only one phase transition and the other(s) are ignored. Strictly speaking, this leads to thermodynamically unstable solutions (violating the second law of thermodynamics), which, however, can be justified by making additional assumptions. We give a more elaborate assessment of reconfinement and multiple phase transitions at the end of this section.

\begin{figure}
    \includegraphics[width = 1. \columnwidth]{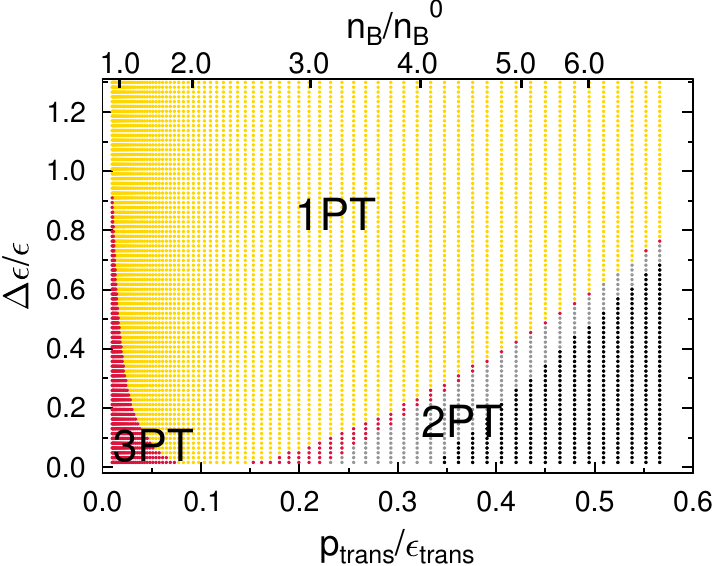}
\caption{\label{fig_PT_study} This figure shows the number of phase transitions that occur for given $\Delta \epsilon$ and $p_{trans}$. The yellow dots represent cases with one phase transition (HQ), red dots with three phase transitions (HQHQ), and grey dots with two phase transitions (QHQ), where quark matter is absolutely stable. The black dots represent QHQ cases, where even negative energy densities occur.}
\end{figure}

Considering the parameter range of Fig.~\ref{4Alfordcases}, we find that there are one, two, or three phase transitions possible, as shown in Fig.~\ref{fig_PT_study}.
The yellow dots in Fig.~\ref{fig_PT_study} represent the cases where only one phase transition happens. It occurs from hadronic to quark matter (HQ transition) and therefore does not differ from the transition points chosen manually in the original parameter scan. 
The red dots correspond to cases with three phase transitions (HQHQ), as discussed for Fig.~\ref{fig_p_mu_3PT_fullrange}. 
The grey dots describe cases with two phase transitions (QHQ). They differ from the first two, since quark matter exists also at the lowest densities. At intermediate densities reconfinement happens, a phase with hadronic matter appears, which disappears again in a deconfinement transition at higher densities. The resulting compact stars of QHQ cannot be considered as hybrid stars in a classical sense, but more as quark stars with a thin hadronic shell somewhere in their interior. 
In fact, as quark matter is the ground state at lowest densities, this case corresponds to absolutely stable strange quark matter. The black dots represent unphysical cases, where, on top of that, quark matter even has negative energy densities. For these reasons we will not consider the QHQ cases as viable models for the SN EOS in our hybrid star analysis.

\begin{figure}
    \includegraphics[width = 1. \columnwidth]{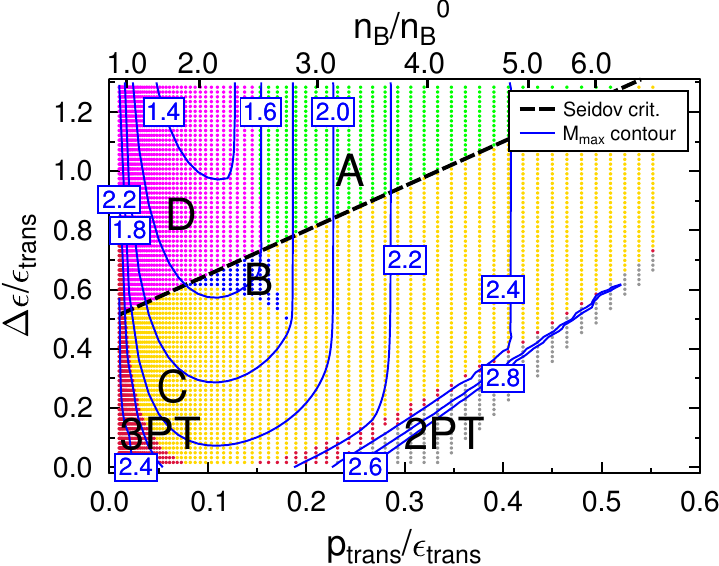}
\caption{\label{fig_Multiple_PT} 
Results for the parameter scan taking into account the occurrence of multiple phase transitions. Red dots show parameter combinations resulting in three phase transitions (HQHQ), grey in two phase transitions (QHQ) and absolutely stable strange quark matter. The other points have only one phase transition (HQ), for which the Alford classification (A, B, C, or D) can be done. The empty triangular region in the lower right corner corresponds to unphysical EOSs with negative energy density, for which the $M$-$R$ relation has not been calculated. 
}
\end{figure}

Figure \ref{fig_Multiple_PT} shows the parameter scan  taking multiple phase transitions into account. In addition to the color coding used in Fig.~\ref{4Alfordcases}, which distinguishes the type of hybrid star, A, B, C, and D, red dots show cases with a reconfinement transition (HQHQ), and the grey dots cases of QHQ where strange quark matter is absolutely stable. For such cases with multiple phase transitions, the hybrid-star classification of Alford cannot be applied.

The maximum mass contour lines up to 2 M$_\odot$ and slightly above lie completely in the HQ area. Therefore, they correspond exactly to the ones shown in Fig.~\ref{4Alfordcases}. This is also true for the regions of hybrid star cases A and B. Imposing strict thermodynamic stability has a strong effect on the maximum masses in the other regions, which would have been cases D or C otherwise. With strict thermodynamic stability masses above the maximum mass of the HS(DD2) EOS are not possible anymore in the lower left corner. For example, for the phase transition parameters which are used in Fig.~\ref{fig_p_mu_3PT_fullrange} ($p_{\rm trans}/\epsilon_{\rm trans} \simeq 0.014$ and $\Delta \epsilon/\epsilon_{\rm trans}=0.2$) and which are situated in the three phase transition region, the maximum mass changes from 2.53~M$_\odot$ (HQ) to 2.42~M$_\odot$ (HQHQ). 
QHQ phase transitions on the other hand, with the additional occurrence of quark matter at low densities taken into account, lead to an increased maximum mass above the one of HS(DD2). This is due to the dominance of quark matter in these compact stars, which in fact are almost pure quark stars. The triangular region in the lower right without points represents the unphysical cases where negative energy densities would occur, for which the $M$-$R$ relations are not calculated.
Finally we note that the phase transition parameters extracted for the hybrid EOSs of Table~\ref{tab_sagert}, and which are shown in Fig.~\ref{search5}, do not lead to the problem of reconfinement, at least not for the HS(DD2) hadronic EOS employed in the present study. The same is true for the phase transition parameters belonging to the example of Fig.~\ref{fig_ms100a4148}.

Chamel \textit{et al}.\ also discussed the possibility of multiple phase transitions for a few example EOSs \cite{2013AA...553A..22C}. They observed the same behavior as discussed above in the HQHQ case: a first phase transition to quark matter, then another one back to hadronic matter, and finally a last one to quark matter with increasing pressure is observed. It is also mentioned that the appearance of quark deconfinement in the strictly thermodynamically stable setup always leads to a lowering of the maximum mass. In contrast, if only one phase transition is enforced, the maximum mass can be increased, as we observe too.

Similar results were obtained by Zdunik and Haensel \cite{zdunik13}. They showed that the reconversion of quark matter back to hadronic matter limits the size of the quark core in their hybrid stars. The resulting maximum mass of the hybrid star has almost the same value as the neutron star consisting of pure hadronic matter when thermodynamic stability is taken into account. Only by ignoring reconfinement can an increased maximum mass be obtained.

The occurrence of reconfinement and multiple phase transitions should probably not be taken as a physically realistic scenario, but rather as artifacts of the EOS models.
The purpose of our investigation is just to show for which phase transition parameters they occur. If several phase transitions are present, this could be taken as an indication that the quark EOS parameters are unrealistic. 
One can also argue that the hadronic EOS is not appropriate at high densities and neither is the quark EOS at low densities, if one uses a two-phase approach as in the present study; see also \cite{zdunik13,blaschke16}. The hadronic EOS model does not account for chiral symmetry restoration and deconfinement, whereas the quark EOS model usually does not account for confinement and the saturation properties of nuclear matter. 
In \cite{zdunik13,blaschke16} it was discussed that the problem of reconfinement could be cured by taking into account the finite size of baryons in the hadronic EOS. From this perspective it is acceptable to enforce just one phase transition and ignore the others as is done in the original parameter scans of Fig.~\ref{4Alfordcases} and \cite{PhysRevD.88.083013,2015PhRvD..92d5022Z,2015PhRvD..92d5022Z,Zacchi16,burgio16,ranea16,2015PhRvD..92h3002A}.


\section{Hyperons}
\label{sec_hyperons}
\begin{figure}
    \includegraphics[width = 1. \columnwidth]{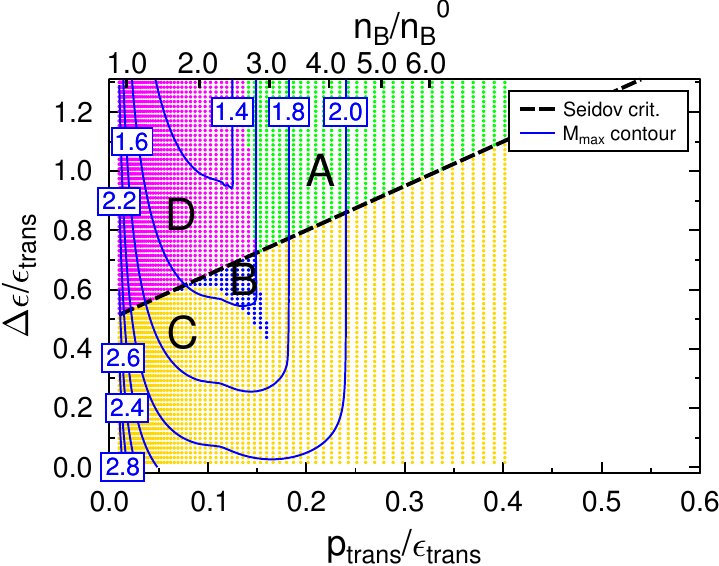}
\caption{\label{fig_BHB_Alford} As in Fig.~\ref{4Alfordcases}, but including hyperons by using the BHB$\Lambda \phi$ EOS instead of HS(DD2).}
\end{figure}

With increasing density, hyperons such as $\Lambda$'s and $\Xi^{-}$'s can appear.
To investigate their effect on the maximum mass of hybrid stars, Fig.~\ref{fig_BHB_Alford} shows a parameter scan using the  BHB$\Lambda \phi$ EOS (see Sec.~\ref{sec_modeling}) for the hadronic part. 
We remind the reader that BHB$\Lambda \phi$ is an extension of HS(DD2) where only $\Lambda$ hyperons have been added. Thus it is identical to HS(DD2) at low densities and temperatures. 
For BHB$\Lambda \phi$ we calculate hybrid stars only up to $p_{\rm trans}/\epsilon_{\rm trans} \approx 0.4$, which is the highest value available in this EOS table.
Comparing Fig.~\ref{fig_BHB_Alford} with the previous parameter scan using the nucleonic HS(DD2) EOS shown in Fig.~\ref{4Alfordcases}, we find no qualitative difference in the distribution of the four different hybrid star cases. Since the $\Lambda$ hyperons appear at around $p/\epsilon = 0.11$ a slight kink in the maximum mass contour lines is visible there. For phase transitions at lower pressures, the results of Fig.~\ref{fig_BHB_Alford} are identical to those of Fig.~\ref{4Alfordcases} because hyperons are not yet present. 
At higher phase transition pressures, the maximum masses are generally reduced due to the presence of hyperons in the hadronic part of the hybrid star. For $p_{\rm trans}/\epsilon_{\rm trans}>0.34$, the phase transition pressure is above the central pressure of the heaviest stable hadronic star, so that quark matter does not appear in stable compact stars and the results are identical to the purely hadronic calculations. We remark that the part of the parameter space we are interested in (case D at low transition pressures), and also our example hybrid EOS used in Fig.~\ref{fig_ms100a4148}, is not affected from the presence of hyperons.

\begin{figure}
    \includegraphics[width = 1. \columnwidth]{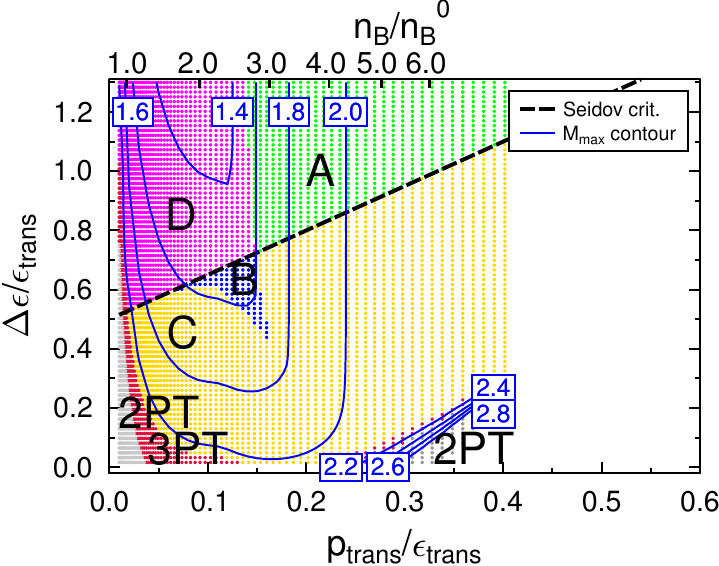}
\caption{\label{fig_BHB_Multiple_PT} As in Fig.~\ref{fig_Multiple_PT}, but including hyperons by using the BHB$\Lambda \phi$ EOS instead of HS(DD2).}
\end{figure}

As discussed in Sec.~\ref{sec_reconfinement} for HS(DD2), strict thermodynamic stability can lead to the appearance of multiple phase transitions. In Fig.~\ref{fig_BHB_Multiple_PT} we repeat the parameter scan for BHB$\Lambda \phi$ but this time taking into account strict thermodynamic stability. As in Fig.~\ref{fig_Multiple_PT} one, two and three phase transitions are possible. The red dots indicate again the cases with three phase transitions (HQHQ). At low $\Delta \epsilon / \epsilon_{\rm trans}$, the region of three phase transitions expands up to $p_{\rm trans}/\epsilon_{\rm trans} = 0.125$ and appears again at around $p_{\rm trans}/\epsilon_{\rm trans} = 0.22$. Compared to Fig.~\ref{fig_Multiple_PT}, the region is shifted to slightly higher values.
The cases with two phase transitions represented by grey dots appear in two regions: at high $p_{\rm trans} / \epsilon_{\rm trans}$, these QHQ cases are the same as in Fig.~\ref{fig_Multiple_PT}. These quark stars show again masses well above the maximum mass of the BHB$\Lambda \phi$ EOS.
When using the BHB$\Lambda \phi$ EOS, an additional two phase transition region appears at low $p_{\rm trans}/\epsilon_{\rm trans}$. Whereas with the HS(DD2) EOS, this region was populated by HQHQ cases, now only an HQH sequence of phase transitions happens. At high densities, the hadronic EOS including hyperons remains favored over quark matter. The maximum masses of these cases are still below the maximum mass of the  BHB$\Lambda \phi$ EOS, as we observed for the strictly thermodynamically stable parameter scan for the HS(DD2) EOS. 
This is similar to the results of \cite{zdunik13}. Without considering reconfinement, it was found that the phase transition to quark matter can resolve the hyperon puzzle; i.e., it can increase the too low maximum mass of a hyperonic EOS to sufficiently high values. If reconfinement is permitted, the maximum mass of the hybrid EOS remains very similar or becomes lower than the one of the hadronic EOS.

\section{Summary and Conclusions}
\label{sec_summary}
The main aim of the present study is to systematically explore the quark matter parameter space for a state-of-the-art hadronic EOS in order to generate a new hybrid SN EOS in the future. All of the currently existing hybrid SN EOS employ STOS for the hadronic part, which is known to have an unrealistically high symmetry energy, and only two of all of the models are compatible with the 2~M$_\odot$ constraint. Here we chose HS(DD2), which has good nuclear matter properties, for the hadronic part. Quark matter is described  
by a constant speed of sound (CSS) EOS with $c_s^2 = 1/3$.
Using these two EOSs we perform a parameter scan as introduced by Alford et al.\ \cite{PhysRevD.88.083013}, where the phase transition density and energy density discontinuity are systematically varied. 
In addition to analyzing the maximum mass, we also apply Alford's classification scheme, resulting in four different types of hybrid stars.
Overall the results look similar as in \cite{PhysRevD.88.083013} where different hadronic EOSs were used: we find the same qualitative distribution of the four hybrid star categories.

We showed that the simple CSS parametrization for quark matter is equivalent to the thermodynamic bag model with $m_s=0$ and an additional term from strong interactions that scales with $\mu^4$. This identification is quite important for our purposes, as the CSS parametrization does not provide a temperature and composition dependence required for CCSN simulations.
$m_s=0$ is not considered as a realistic value and often $m_s = 100$~MeV is used instead. 
A finite strange quark mass induces a nonconstant speed of sound, and thus lifts the one-to-one correspondence between the CSS and the thermodynamic bag model EOSs. It also changes the resulting phase transition properties $p_{\rm trans}/\epsilon_{\rm trans}$ and $\Delta \epsilon / \epsilon_{\rm trans}$. However, we showed that for $m_s=100$~MeV the speed of sound shows only little deviation from the fixed value $c_s^2 = 1/3$.

In order to get insights about the quark matter parameter regions which are favorable for CCSN explosions, we calculated $p_{\rm trans}/\epsilon_{\rm trans}$ and $\Delta \epsilon / \epsilon_{\rm trans}$ of the already existing hybrid EOS from Sagert \textit{et al}.\ and Nakazato \textit{et al}., and added this information to the parameter scan.
The EOSs that showed explosions in one-dimensional CCSN simulations are all situated in the parameter region that leads to a disconnected third family of compact stars. This supports our initial considerations that third-family features might play an important role in the CCSN explosion mechanism induced by the QCD phase transition; see also \cite{2015arXiv151106551H} for further details.

Regarding the question whether this mechanism can still work despite the 2~M$_\odot$ constraint the results do not look very promising at first. To form a third family in cold compact stars requires phase transition densities below 2.5~$n_B^0$. On the other hand, to reach sufficiently high maximum masses, the energy density discontinuity has to be rather low, meaning that the phase transition is rather weak and the third family is not very pronounced. In consequence, only a small parameter region remains where one has a third family and a maximum mass above 2~M$_\odot$. From this region we presented the $M$-$R$ relation of one potential future hybrid SN EOS, employing the bag model parameters $\alpha_s = 0.7$ and $B^{1/4} = 138.5$ MeV.
The energy density discontinuity of our example case is lower than the one of B139 which was not (yet) found to explode. However, we want to emphasize again that the existing hybrid EOSs have been tested only for very few progenitor models. It is not excluded that even a slightly more pessimistic EOS still could trigger explosions for other progenitors. The effect of the hadronic EOS and of local vs.\ global charge neutrality for the phase transition also remains to be studied.

Considering hyperons in the hadronic EOS using the BHB$\Lambda \phi$ EOS did not show a qualitative difference in the distribution of the four Alford cases.  Again only a small parameter region remained which might be interesting for future SN EOS candidates. Since the transition pressures in this region lie below the pressure where hyperons appear, our proposed example case for a new hybrid SN EOS would not be affected by the presence of hyperons.

The penultimate part of our paper dealt with the reconfinement problem. 
Using the assumption of chemical equilibrium at the transition point, the pressure-baryon chemical potential relation of the CSS EOS can be derived.
The $p(\mu_B)$ relations revealed that multiple 
phase transitions are possible in our considered range of the parameter scan. 
Three cases were identified: one (hadron-quark, HQ); two (quark-hadron-quark, QHQ); or three (hadron-quark-hadron-quark, HQHQ) phase transitions, where the second case corresponds to a special form of absolutely stable strange quark matter and in the third case a spurious reconfinement and subsequent second deconfinement transitions occur.
Low $p_{\rm trans}$ and small $\Delta \epsilon$ (in regions which otherwise belong to cases C and D) lead to HQHQ, whereas high $p_{\rm trans}$ and low $\Delta \epsilon$ (in regions which otherwise belong to case C) lead to QHQ and corresponding strange quark star configurations. 
For BHB$\Lambda \phi$ an additional two-phase-transition case (hadron-quark-hadron, HQH) at low transition pressures is present.
In Sec.~\ref{sec_reconfinement} we discussed different options for how to interpret and deal with reconfinement. If it occurs in a density region where one of the two EOSs is not reliable any more, it is justified to ignore it. Otherwise it might point to a region of the quark matter parameter space which is not realistic and should be avoided.

Without considering multiple phase transitions, i.e., ignoring thermodynamic stability in a strict sense, we found that the hybrid stars can have a maximum mass above the ones of the purely hadronic EOSs HS(DD2) and BHB$\Lambda \phi$. Conversely, if strict thermodynamic stability is taken into account, this is not possible, unless one has absolutely stable quark matter. \citet{masuda13,masuda16} constructed a crossover phase change by an interpolation between the hadronic and the quark EOS instead of using the usual Maxwell or Gibbs construction. In this procedure, which represents a ``manual'' manipulation of the EOS, the maximum mass can be increased. One can conclude that without further assumptions the inclusion of quark matter (using $c_s^{2}=1/3$) generally leads to a reduction of the maximum mass. Only by making use of additional assumptions (e.g., crossover or suppression of reconfinement) the maximum mass can be increased.
It would be interesting to further explore the role of multiple phase transitions for other hadronic EOSs in the context of the hyperon puzzle (similarly as in \cite{zdunik13}), especially as HS(DD2), and to a smaller extent also BHB$\Lambda \phi$, are particularly stiff EOSs at high densities. 

We conclude that suitable parameters for a new hybrid SN EOS have to be searched for in a strongly restricted region of $\Delta \epsilon$ and $p_{\rm trans}$, where a maximum mass of 2 M$_\odot$ is obtained and a third-family feature is found. Additionally, the possibility of multiple phase transitions also has to be considered.
In future work, the parameter scan should be repeated with speed of sound values above $c_s^2=1/3$. An increase of $c_s^2$ leads to stronger third-family features in the $M$-$R$ relation and generally higher maximum masses \cite{PhysRevD.88.083013,2016EPJA...52...62A,2013arXiv1310.3803B,ranea16} and thus is favorable for realizing the CCSN explosion mechanism induced by the QCD phase transition.
A speed of sound above $c_s^2=1/3$, can, e.g., be realized by introducing vector interactions; see \cite{Benic2014,Zacchi16,ranea16,klaehn15,klaehn16}.
Additionally, the influence of extra interaction parameters such as $a_2$ might be analyzed.
It would also be interesting to compare with the high-density limit of perturbative QCD \cite{2010PhRvD..81j5021K,kurkela16}.
Currently we are calculating a new hybrid SN EOS using the parameters from the example case shown in this paper. Once ready, we will test it in one-dimensional CCSN simulations for several progenitors and eventually in multidimensional simulations to investigate the effects of the hadron-quark phase transition in CCSNe and whether it still can lead to explosions.

\section{Acknowledgements}
The authors acknowledge useful discussions and input from M.\ Liebend\"orfer, M.~Alford, and S.~Han. M.H. would like to thank A.\ Drago and F.\ Burgio for stimulating discussions and useful comments. 
This work has been supported by the European Research Council (Framework Program 7) under European Research Council Advanced Grant Agreement No. 321263 - FISH and the Swiss National Science Foundation.

\bibliographystyle{apsrev}
\bibliography{literat_here}

\end{document}